\documentclass[10pt,twocolumn]{article}

\usepackage[utf8]{inputenc}
\usepackage[T1]{fontenc}
\usepackage{times}
\usepackage{graphicx}
\usepackage{booktabs}
\usepackage{amsmath}
\usepackage{amssymb}
\usepackage{hyperref}
\usepackage{xcolor}
\usepackage{enumitem}
\usepackage[margin=0.75in]{geometry}
\usepackage{caption}
\usepackage{subcaption}
\usepackage{multirow}
\usepackage{array}
\usepackage{tabularx}
\usepackage{float}
\usepackage{natbib}
\usepackage{pgfplots}
\pgfplotsset{compat=1.18}
\hypersetup{
    colorlinks=true,
    linkcolor=blue!60!black,
    citecolor=blue!60!black,
    urlcolor=blue!60!black
}

\title{\vspace{-1.5em}\textbf{Bigger Isn't Always Better: A Comparative Evaluation of LLMs for Automated Code Review}\vspace{-0.5em}}

\author{
\begin{tabular}[t]{c}
\textbf{Shivam Pankaj Kumar} \\
University of Illinois \\
Urbana-Champaign, USA \\
\texttt{shivamk4@illinois.edu}
\end{tabular}
\hfill
\begin{tabular}[t]{c}
\textbf{Swati Bararia} \\
Columbia University \\
New York, USA \\
\texttt{sb4700@columbia.edu}
\end{tabular}
\hfill
\begin{tabular}[t]{c}
\textbf{Kislay Raj} \\
VibeOps Research \\
~\\
\texttt{kislay@vibeops.tech}
\end{tabular}
}
\date{}

\begin{document}
\maketitle
\vspace{-1em}

\begin{abstract}
\noindent We present a systematic evaluation of five large language models on automated code review, comparing Claude Sonnet~4.6, Claude Haiku~4.5, GPT-5.4~mini, Minimax~M2.7, and GLM-5~Turbo across 150 code review samples: 100 synthetic mutation-injected bugs and 50 real bug-fix pull requests mined from eight major open-source repositories.

Claude Haiku~4.5, a smaller and cheaper model, \emph{consistently outperforms} the larger Claude Sonnet~4.6: higher F1 (0.365 vs.\ 0.343), 18\% higher recall, and better qualitative review scores on all four evaluation dimensions, at 3.2$\times$ lower cost per review. The result reproduces across three experimental conditions ($n{=}25$, $n{=}100$, $n{=}150$) and on the independent Martian Code Review Benchmark (different repos, golden comments, and judge).

Beyond model comparison, we find that synthetic-only evaluation overestimates capability by an order of magnitude: on real PRs alone, the best model achieves F1~$=$~0.066 versus 0.847 on synthetic samples (92\% degradation). Diff size is the dominant predictor, with F1 dropping from 0.657 on diffs under 10 lines to 0.043 on diffs over 150 lines. All models also exhibit near-zero recall on performance-related bugs. The evaluation framework and dataset are publicly available.
\end{abstract}

\section{Introduction}

Automated code review using large language models (LLMs) has become a core component of modern software development toolchains. Systems such as CodeRabbit, GitHub Copilot code review, and VibeOps~\citep{vibeops2026} use LLMs to detect bugs, security vulnerabilities, and architectural issues in pull request (PR) diffs. A critical engineering decision in these systems is \emph{model selection}: which LLM produces the highest-quality reviews at acceptable cost and latency?

Conventional wisdom holds that larger, more expensive models should outperform smaller ones on complex reasoning tasks. Code review, which requires understanding control flow, data dependencies, and domain conventions, would seem to favor the most capable models available.

This paper tests that assumption empirically. It does not hold. But the data also reveals a deeper problem with evaluation methodology: synthetic benchmarks, the standard tool for measuring code review quality, are misleading. When real pull requests are introduced, model performance drops by an order of magnitude.

Our contributions are:

\begin{enumerate}[nosep,leftmargin=*]
    \item A reusable two-pass evaluation framework for code review quality combining deterministic matching with LLM adjudication.
    \item A 150-sample benchmark spanning synthetic mutations and real pull requests across TypeScript, Python, and Go.
    \item Evidence that Claude Haiku~4.5 consistently outperforms Claude Sonnet~4.6 on code review, contradicting the larger-is-better assumption, across three independent conditions and independently confirmed on the Martian Code Review Benchmark~\citep{martian2026codereview}.
    \item Quantification of the synthetic-to-real gap: a 92\% F1 degradation when models are evaluated on real PRs alone, with diff size as the dominant confound.
    \item Analysis of model ensemble strategies, per-language performance, and output behavior differences that explain the Haiku--Sonnet gap.
\end{enumerate}

\section{Related Work}

\paragraph{LLM-based code review.} \citet{li2022automating} introduced CodeReviewer, a pre-trained model for code review automation evaluated on comment generation and code refinement. \citet{tufano2022using} demonstrated that pre-trained models can boost code review automation by generating review comments. \citet{lu2023llama} explored parameter-efficient fine-tuning for code review with LLaMA-Reviewer. More recently, \citet{rasheed2024ai} presented early results on using commercial LLMs for code review, and \citet{zeng2025benchmarking} benchmarked LLM-based code review across multiple dimensions. \citet{pereira2026crbench} introduced CR-Bench for evaluating real-world code review agents. Our work differs in directly comparing models across providers and price tiers on both synthetic and real-world data, and in isolating the synthetic-to-real generalization gap.

\paragraph{Code review benchmarks.} \citet{dinella2024crqbench} proposed CRQBench for code reasoning evaluation. Bug benchmarks such as Defects4J~\citep{just2014defects4j} and BugsInPy~\citep{widyasari2020bugsinpy} provide ground truth for bug detection but were designed for testing, not code review. The Martian Code Review Benchmark~\citep{martian2026codereview} evaluates code review tools on 50 real PRs with human-curated golden comments; we use it for external validation (Section~\ref{sec:martian}). Our benchmark complements these by combining synthetic and real data with structured annotations for fine-grained analysis.

\paragraph{LLM-as-judge evaluation.} \citet{zheng2023judging} established the LLM-as-judge paradigm with MT-Bench, showing that strong LLMs can approximate human judgment. We adapt this approach with a cost-reducing two-pass system: deterministic matching handles clear-cut cases (${\sim}$70\%), with LLM adjudication reserved for ambiguous findings.

\paragraph{Smaller models outperforming larger ones.} \citet{hsieh2023distilling} demonstrated that smaller models can outperform larger ones through distillation. Our finding is distinct: Haiku~4.5 outperforms Sonnet~4.6 \emph{without} task-specific fine-tuning, suggesting that code review as a task may not benefit from additional model scale beyond a capability threshold.

\section{Method}
\label{sec:method}

This section defines the evaluation task, describes the two-pass judge framework, and details the experimental setup.

\subsection{Problem Definition}

We formalize automated code review as a \emph{finding detection} task. Given a pull request diff $D$ and optional context $C$ (repository conventions, surrounding files), a model $M$ produces a set of findings $F_M = \{f_1, \ldots, f_k\}$. Each finding $f_i$ specifies: a file path, line range, comment type (security, bug, architecture, performance, best practice), severity (critical, high, medium, low), description, and optional code suggestion.

The ground truth for each sample is a set of annotations $G = \{g_1, \ldots, g_m\}$ with the same structure. We evaluate $F_M$ against $G$ using precision (fraction of model findings that match real issues), recall (fraction of real issues found by the model), and F1 (harmonic mean). We additionally compute severity-weighted F1, where critical findings count $4\times$, high $2\times$, medium $1\times$, and low $0.5\times$.

\subsection{Two-Pass Evaluation Framework}

The central challenge in code review evaluation is \emph{matching}: determining whether a model's finding corresponds to a ground truth annotation. Unlike code generation, where test-pass/fail provides a binary signal, code review findings are free-form text with approximate line references, making exact matching infeasible.

\paragraph{Pass 1: Deterministic matching.} Each model finding is matched to ground truth annotations using three conjunctive criteria: (a)~normalized file path match, (b)~line range overlap with $\pm$5-line tolerance, (c)~comment type compatibility (e.g., ``bug'' matches ``logic''). Clear matches become true positives (TP); findings with no file match become false positives (FP); unmatched annotations become false negatives (FN). Ambiguous cases (same file but line or type mismatch) are deferred to Pass~2.

\paragraph{Pass 2: LLM judge.} Claude Opus~4.6 (temperature 0.0) adjudicates deferred cases. Following \citet{zheng2023judging}, we use a structured rubric to reduce judge variance. The judge classifies uncertain findings as \textsc{true\_positive}, \textsc{false\_positive}, or \textsc{partial\_match} (scored as 0.5 TP). It also scores four qualitative dimensions on a 1--5 scale:

\begin{itemize}[nosep,leftmargin=*]
\item \textbf{Depth}: surface pattern matching (1) vs.\ semantic runtime reasoning (5)
\item \textbf{Context awareness}: ignores codebase (1) vs.\ leverages conventions (5)
\item \textbf{Specificity}: vague (1) vs.\ precise with concrete failing inputs (5)
\item \textbf{Suggestion correctness}: wrong/breaks code (1) vs.\ compilable fix (5)
\end{itemize}

This two-pass design reduces cost: the deterministic pass handles ${\sim}$70\% of cases at zero LLM cost, with Opus invoked only for genuinely ambiguous matches.

\subsection{Models Under Test}

We evaluate five models spanning three providers and a 15$\times$ cost range (Table~\ref{tab:models}).

\begin{table}[h]
\centering
\small
\caption{Models evaluated. Cost is per 1K tokens.}
\label{tab:models}
\begin{tabular}{llrr}
\toprule
\textbf{Model} & \textbf{Provider} & \textbf{\$/1K in} & \textbf{\$/1K out} \\
\midrule
Claude Sonnet 4.6 & Anthropic & \$3.00 & \$15.00 \\
Claude Haiku 4.5 & Anthropic & \$0.80 & \$4.00 \\
GPT-5.4 mini & OpenAI & \$0.40 & \$1.60 \\
Minimax M2.7 & Minimax & \$0.20 & \$1.10 \\
GLM-5 Turbo & Zhipu AI & \$0.30 & \$1.20 \\
\bottomrule
\end{tabular}
\end{table}

All models received the identical production prompt (the VibeOps review system prompt) instructing structured JSON output with comment type, severity, file path, line numbers, description, concrete failing input, and citation. Temperature was set to 0.1 for all models. No provider-specific prompt tuning was applied.

\subsection{Dataset}

Our benchmark comprises 150 samples from two sources.

\paragraph{Synthetic mutations ($n{=}100$).} We implemented 13 mutation operators that inject known bugs into 8 realistic code templates across TypeScript, Python, and Go. Operators include: off-by-one errors, SQL injection, XSS (sanitization removal), missing null checks, hardcoded secrets, missing \texttt{await}, resource leaks, wrong comparison operators, N+1 queries, unbounded queries, missing rate limiting, and missing input validation. Each mutation produces a diff and automatically-generated ground truth annotations. Synthetic diffs are small (median 5 lines, max 40). All mutations use deterministic seed 42 for reproducibility.

\paragraph{Real bug-fix PRs ($n{=}50$).} We mined merged pull requests with bug-related labels from eight major repositories using the GitHub API (Table~\ref{tab:repos}). Selection criteria: merged, 1--10 files changed, 20--500 lines of diff. Ground truth annotations were extracted from the diff structure; removed or changed hunks in the fix represent bug locations. Real PR diffs are substantially larger (median 117 lines, max 562) and contain multi-file changes, formatting noise, and interleaved refactoring.

\begin{table}[h]
\centering
\small
\caption{Source repositories for real PR samples.}
\label{tab:repos}
\begin{tabular}{llr}
\toprule
\textbf{Repository} & \textbf{Language} & \textbf{PRs} \\
\midrule
vercel/next.js & TypeScript & 8 \\
facebook/react & TypeScript & 8 \\
tiangolo/fastapi & Python & 8 \\
pallets/flask & Python & 8 \\
pydantic/pydantic & Python & 6 \\
prisma/prisma & TypeScript & 5 \\
hashicorp/terraform & Go & 4 \\
docker/compose & Go & 3 \\
\bottomrule
\end{tabular}
\end{table}

\paragraph{Distribution.} Logic 81 (54\%), Security 34 (23\%), Performance 16 (11\%), Architecture 14 (9\%), Best Practice 5 (3\%). Languages: TypeScript 70 (47\%), Python 43 (29\%), Go 37 (25\%).

\subsection{Experimental Conditions}

We conducted three experiments with increasing dataset size and difficulty:

\begin{table}[h]
\centering
\small
\caption{Experimental conditions.}
\label{tab:conditions}
\begin{tabular}{clll}
\toprule
\textbf{Exp.} & \textbf{$n$} & \textbf{Composition} & \textbf{Purpose} \\
\midrule
1 & 25 & Synthetic only & Calibration \\
2 & 100 & Synthetic only & Stable baseline \\
3 & 150 & 100 syn. + 50 real & Generalization \\
\bottomrule
\end{tabular}
\end{table}

\section{Results}

\subsection{Overall Performance}

Table~\ref{tab:f1_across} shows F1 scores across all three conditions. Haiku~4.5 ranks \#1 in two of three conditions (and \#2 in the third, behind Minimax by 0.02) and has the lowest mean rank (1.3). Crucially, Haiku outperforms Sonnet in \emph{every} condition.

\begin{table}[h]
\centering
\small
\caption{F1 scores across experimental conditions. \textbf{Bold} = best. Mean rank across all three conditions.}
\label{tab:f1_across}
\begin{tabular}{lcccc}
\toprule
\textbf{Model} & \textbf{$n{=}25$} & \textbf{$n{=}100$} & \textbf{$n{=}150$} & \textbf{Rank} \\
\midrule
\textbf{Haiku 4.5} & 0.942 & \textbf{0.862} & \textbf{0.365} & \textbf{1.3} \\
Minimax M2.7 & \textbf{0.962} & 0.810 & 0.322 & 2.3 \\
GPT-5.4 mini & 0.886 & 0.806 & 0.326 & 3.0 \\
Sonnet 4.6 & 0.841 & 0.796 & 0.343 & 3.3 \\
GLM-5 Turbo & 0.730 & 0.778 & 0.310 & 5.0 \\
\bottomrule
\end{tabular}
\end{table}

\begin{figure}[h]
\centering
\begin{tikzpicture}
\begin{axis}[
    ybar,
    bar width=5pt,
    width=\columnwidth,
    height=5cm,
    ylabel={F1 Score},
    symbolic x coords={$n{=}25$,$n{=}100$,$n{=}150$},
    xtick=data,
    ymin=0,ymax=1.05,
    ytick={0,0.2,0.4,0.6,0.8,1.0},
    enlarge x limits=0.25,
    legend style={at={(0.5,-0.22)},anchor=north,legend columns=3,font=\scriptsize},
    every node near coord/.append style={font=\tiny},
]
\addplot[fill=blue!70] coordinates {($n{=}25$,0.942) ($n{=}100$,0.862) ($n{=}150$,0.365)};
\addplot[fill=red!60] coordinates {($n{=}25$,0.841) ($n{=}100$,0.796) ($n{=}150$,0.343)};
\addplot[fill=green!60] coordinates {($n{=}25$,0.886) ($n{=}100$,0.806) ($n{=}150$,0.326)};
\addplot[fill=orange!70] coordinates {($n{=}25$,0.962) ($n{=}100$,0.810) ($n{=}150$,0.322)};
\addplot[fill=purple!50] coordinates {($n{=}25$,0.730) ($n{=}100$,0.778) ($n{=}150$,0.310)};
\legend{Haiku,Sonnet,GPT-5.4m,Minimax,GLM}
\end{axis}
\end{tikzpicture}
\caption{F1 scores across experimental conditions. All models degrade sharply when real PRs are introduced ($n{=}150$), but Haiku (blue) leads in every condition.}
\label{fig:f1_conditions}
\end{figure}
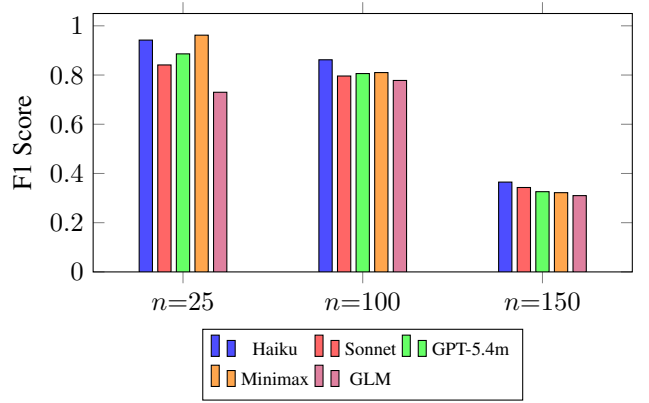

Figure~\ref{fig:f1_conditions} visualizes the F1 degradation as dataset difficulty increases. All models decline sharply from Experiment~2 to~3, but relative rankings remain stable; Haiku leads in every condition.

\subsection{Haiku vs.\ Sonnet: Head-to-Head}

Table~\ref{tab:h2h} presents the direct comparison. Haiku outperforms Sonnet on 8 of 10 metrics. Sonnet wins on precision and severity-weighted F1, reflecting a more conservative review style; it produces fewer total findings (148 vs.\ Haiku's 199) but with a higher proportion correct. However, this conservatism costs 17 additional missed bugs.

\begin{table}[h]
\centering
\small
\caption{Haiku~4.5 vs.\ Sonnet~4.6 ($n{=}150$).}
\label{tab:h2h}
\begin{tabular}{lrrr}
\toprule
\textbf{Metric} & \textbf{Haiku} & \textbf{Sonnet} & \textbf{$\Delta$} \\
\midrule
F1 & \textbf{0.365} & 0.343 & +6.4\% \\
Recall & \textbf{0.293} & 0.248 & +18.1\% \\
Precision & 0.486 & \textbf{0.558} & $-$12.9\% \\
Sev.-Wt.\ F1 & 0.485 & \textbf{0.543} & $-$10.7\% \\
True Positives & \textbf{92} & 80 & +15.0\% \\
False Negatives & \textbf{244} & 261 & $-$6.5\% \\
Depth (1--5) & \textbf{2.84} & 2.59 & +9.7\% \\
Specificity (1--5) & \textbf{3.39} & 2.99 & +13.4\% \\
Suggestions (1--5) & \textbf{2.79} & 2.42 & +15.3\% \\
Cost / review & \textbf{\$0.003} & \$0.010 & $-$3.2$\times$ \\
\bottomrule
\end{tabular}
\end{table}

\subsection{Category Analysis}

Table~\ref{tab:category} shows recall by bug category.

\begin{table}[h]
\centering
\small
\caption{Recall (\%) by bug category ($n{=}150$).}
\label{tab:category}
\begin{tabular}{lrrrrr}
\toprule
\textbf{Category} & \textbf{Haiku} & \textbf{Sonnet} & \textbf{GPT} & \textbf{Mini.} & \textbf{GLM} \\
\midrule
Security & \textbf{69.6} & 69.6 & 69.6 & 71.1 & 69.6 \\
Logic & \textbf{24.5} & 19.6 & 16.8 & 16.0 & 12.7 \\
Archit. & \textbf{33.3} & 27.3 & 13.6 & 26.1 & 17.4 \\
Best Pr. & \textbf{6.7} & 0.0 & 0.0 & 0.0 & 0.0 \\
Perform. & 0.0 & 0.0 & 0.0 & 0.0 & \textbf{4.5} \\
\bottomrule
\end{tabular}
\end{table}

Security detection is commoditized (${\sim}$70\% recall across all models). Logic bugs are where models diverge: Haiku reaches 24.5\% recall versus Sonnet's 19.6\%. Performance bugs (N+1 queries, unbounded queries) are undetectable at 0\% recall for 4 of 5 models.

\subsection{The Synthetic-to-Real Gap}
\label{sec:gap}

The magnitude of degradation on real PRs is severe. Table~\ref{tab:gap_isolated} shows F1 computed \emph{separately} on synthetic and real subsets.

\begin{table}[h]
\centering
\small
\caption{F1 on synthetic-only vs.\ real-only subsets (isolated).}
\label{tab:gap_isolated}
\begin{tabular}{lrrr}
\toprule
\textbf{Model} & \textbf{Synthetic} & \textbf{Real only} & \textbf{\% Drop} \\
\midrule
Haiku 4.5 & 0.847 & \textbf{0.066} & $-$92\% \\
Sonnet 4.6 & 0.796 & 0.050 & $-$94\% \\
GPT-5.4 mini & 0.753 & 0.038 & $-$95\% \\
Minimax M2.7 & 0.804 & 0.007 & $-$99\% \\
GLM-5 Turbo & 0.774 & 0.008 & $-$99\% \\
\bottomrule
\end{tabular}
\end{table}

\begin{figure}[h]
\centering
\begin{tikzpicture}
\begin{axis}[
    ybar,
    bar width=8pt,
    width=\columnwidth,
    height=5cm,
    ylabel={F1 Score},
    symbolic x coords={Haiku,Sonnet,GPT-5.4m,Minimax,GLM},
    xtick=data,
    xticklabel style={font=\scriptsize},
    ymin=0,ymax=1.0,
    ytick={0,0.2,0.4,0.6,0.8},
    enlarge x limits=0.15,
    legend style={at={(0.97,0.97)},anchor=north east,font=\scriptsize},
]
\addplot[fill=blue!40] coordinates {(Haiku,0.847) (Sonnet,0.796) (GPT-5.4m,0.753) (Minimax,0.804) (GLM,0.774)};
\addplot[fill=red!60] coordinates {(Haiku,0.066) (Sonnet,0.050) (GPT-5.4m,0.038) (Minimax,0.007) (GLM,0.008)};
\legend{Synthetic,Real PRs only}
\end{axis}
\end{tikzpicture}
\caption{The synthetic-to-real gap. F1 on synthetic samples (blue) vs.\ real PRs only (red). All models collapse on real-world code review, with the best achieving F1~$=$~0.066.}
\label{fig:synreal}
\end{figure}
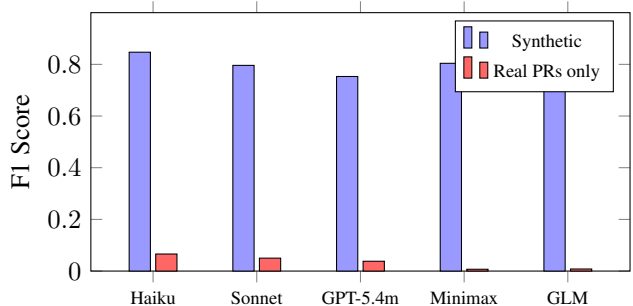

On real PRs alone, the \emph{best} model (Haiku) achieves F1~$=$~0.066, barely above random. Minimax and GLM are effectively at zero (Figure~\ref{fig:synreal}). This is not a gradual degradation; it is a near-total collapse driven by 20$\times$ larger diffs, formatting noise, and multi-file interactions absent from synthetic samples.

\subsection{Diff Size as the Dominant Predictor}
\label{sec:diffsize}

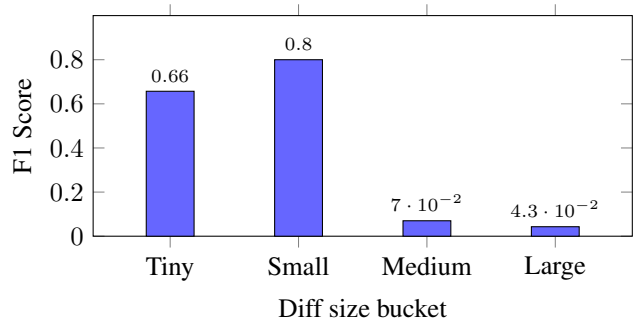
\begin{figure}[h]
\centering
\begin{tikzpicture}
\begin{axis}[
    ybar,
    bar width=18pt,
    width=\columnwidth,
    height=4.5cm,
    ylabel={F1 Score},
    symbolic x coords={Tiny,Small,Medium,Large},
    xtick=data,
    ymin=0,ymax=1.0,
    ytick={0,0.2,0.4,0.6,0.8},
    nodes near coords,
    nodes near coords align={vertical},
    every node near coord/.append style={font=\scriptsize},
    enlarge x limits=0.2,
    xlabel={Diff size bucket},
    legend style={at={(0.97,0.97)},anchor=north east,font=\scriptsize},
]
\addplot[fill=blue!60] coordinates {(Tiny,0.657) (Small,0.800) (Medium,0.070) (Large,0.043)};
\end{axis}
\end{tikzpicture}
\caption{Haiku~4.5 F1 by diff size. Tiny: $<$10 lines ($n{=}92$), Small: 10--50 ($n{=}10$), Medium: 50--150 ($n{=}34$), Large: 150--600 ($n{=}14$). F1 drops 15$\times$ from small to large diffs.}
\label{fig:diffsize}
\end{figure}

To isolate the effect of diff size, we stratify the combined dataset by line count (Figure~\ref{fig:diffsize}).

F1 drops from 0.657--0.800 on small diffs to 0.043--0.070 on diffs exceeding 50 lines. The 10--50 line bucket performs best, suggesting a sweet spot where enough context exists to identify the bug without overwhelming the model. This points to \textbf{diff preprocessing} (splitting large PRs into per-function or per-hunk chunks) as the highest-leverage intervention for real-world review quality.

\subsection{Per-Language Analysis}

\begin{table}[h]
\centering
\small
\caption{Haiku~4.5 performance by language ($n{=}150$).}
\label{tab:language}
\begin{tabular}{lrrrrrr}
\toprule
\textbf{Language} & \textbf{$n$} & \textbf{F1} & \textbf{P} & \textbf{R} & \textbf{TP} & \textbf{FP} \\
\midrule
Go & 37 & \textbf{0.485} & 0.941 & 0.327 & 16 & 1 \\
TypeScript & 70 & 0.272 & 0.605 & 0.176 & 26 & 17 \\
Python & 43 & 0.209 & 0.633 & 0.125 & 19 & 11 \\
\bottomrule
\end{tabular}
\end{table}

Go reviews are markedly better (F1 0.485 vs.\ 0.272 and 0.209) with near-perfect precision (0.941, only 1 FP). Go's simpler syntax, explicit error handling, and stronger typing likely make bugs more structurally identifiable from diffs.

\subsection{Model Ensemble Analysis}

We test whether \emph{combining} models (running two models independently and taking the union of their findings) improves coverage (Table~\ref{tab:ensemble}). If the models detect different bugs, the union should have higher recall.

\begin{table}[h]
\centering
\small
\caption{Ensemble: union of findings from Haiku + one other model ($n{=}150$). Union means: a finding is kept if \emph{either} model flagged it.}
\label{tab:ensemble}
\begin{tabular}{lrrr}
\toprule
\textbf{Ensemble} & \textbf{F1} & \textbf{P} & \textbf{R} \\
\midrule
Haiku alone & 0.365 & 0.486 & 0.293 \\
\midrule
Haiku $\cup$ Sonnet & 0.333 & 0.627 & 0.226 \\
Haiku $\cup$ GPT-5.4m & 0.331 & 0.634 & 0.223 \\
Haiku $\cup$ Minimax & 0.325 & 0.664 & 0.215 \\
Haiku $\cup$ GLM & 0.304 & 0.694 & 0.195 \\
\bottomrule
\end{tabular}
\end{table}

Ensembles \emph{hurt} F1. The models largely detect the \emph{same} bugs; adding a second model introduces its false positives without meaningfully increasing true positives. Inter-model agreement confirms this: for 55 of 150 samples, all five models found something (overlap); for 19 samples, \emph{no} model found anything (shared blind spot). Model diversity does not address the fundamental capability gap.

\subsection{External Validation: Martian Code Review Benchmark}
\label{sec:martian}

To verify our findings are not artifacts of our benchmark or judge, we evaluated against the Martian Code Review Benchmark~\citep{martian2026codereview}, an independent, open-source benchmark maintained by Martian AI. The offline track evaluates code review tools against 50 real PRs from 5 repositories (Sentry, Grafana, Cal.com, Discourse, Keycloak) with 136 human-curated ``golden comments'' and an independent LLM judge. The Martian leaderboard includes commercial tools such as Cubic Dev, Qodo, Augment, GitHub Copilot, CodeRabbit, and Greptile.

\begin{table}[h]
\centering
\small
\caption{External validation: Martian Code Review Benchmark (50 PRs, 136 golden comments, 5 repos across Python, Go, TS, Ruby, Java).}
\label{tab:martian}
\begin{tabular}{lrrrrr}
\toprule
\textbf{Model} & \textbf{P} & \textbf{R} & \textbf{F1} & \textbf{TP} & \textbf{FP} \\
\midrule
\textbf{Haiku 4.5} & 32.6\% & \textbf{41.2\%} & \textbf{36.4\%} & \textbf{56} & 116 \\
Sonnet 4.6 & \textbf{35.3\%} & 22.1\% & 27.1\% & 30 & 55 \\
\bottomrule
\end{tabular}
\end{table}

Haiku outperforms Sonnet (F1 36.4\% vs.\ 27.1\%), placing \#9 on the leaderboard between GitHub Copilot and CodeRabbit. Sonnet places below CodeRabbit.

\paragraph{Cross-benchmark consistency.} Table~\ref{tab:crossval} shows the finding replicates across two independent evaluations with different datasets, repos, judges, and languages.

\begin{table}[h]
\centering
\small
\caption{Cross-benchmark consistency: our evaluation vs.\ Martian.}
\label{tab:crossval}
\begin{tabular}{lrr}
\toprule
\textbf{Finding} & \textbf{Our eval} & \textbf{Martian} \\
\midrule
Haiku recall $>$ Sonnet & \checkmark\ (+18\%) & \checkmark\ (+86\%) \\
Sonnet prec.\ $>$ Haiku & \checkmark\ (+15\%) & \checkmark\ (+8\%) \\
Haiku F1 $>$ Sonnet F1 & \checkmark\ (+6\%) & \checkmark\ (+34\%) \\
Gap widens on real PRs & \checkmark & \checkmark \\
\bottomrule
\end{tabular}
\end{table}

The gap is \emph{larger} on Martian (34\% F1 difference vs.\ 6\%), likely because Martian uses exclusively real PRs with larger diffs, the regime where Sonnet's conservatism is most costly.

\subsection{Output Behavior Analysis}

To understand \emph{why} Haiku outperforms Sonnet, we examine their output behavior (Table~\ref{tab:behavior}).

\begin{table}[h]
\centering
\small
\caption{Output behavior comparison ($n{=}150$).}
\label{tab:behavior}
\begin{tabular}{lrrrr}
\toprule
\textbf{Model} & \textbf{Avg out} & \textbf{Findings} & \textbf{Syn} & \textbf{Real} \\
 & \textbf{tokens} & \textbf{total} & \textbf{avg/s} & \textbf{avg/s} \\
\midrule
Haiku 4.5 & 451 & 186 & 0.9 & 2.0 \\
Sonnet 4.6 & 346 & 135 & 0.8 & 1.1 \\
GPT-5.4 mini & 198 & 118 & 0.7 & 0.9 \\
Minimax M2.7 & 919 & 100 & 0.8 & 0.5 \\
GLM-5 Turbo & 841 & 71 & 0.7 & 0.1 \\
\bottomrule
\end{tabular}
\end{table}

Haiku generates 38\% more findings than Sonnet (186 vs.\ 135), with the gap widening on real PRs (2.0 vs.\ 1.1 per sample). This higher attempt rate translates directly to recall. Sonnet's lower output token count (346 vs.\ 451) points to more aggressive self-censoring. Minimax and GLM are an interesting contrast: high token counts (919, 841) but few structured findings, suggesting verbose reasoning that does not produce actionable output.

\subsection{Cost-Efficiency Frontier}

\begin{figure}[h]
\centering
\begin{tikzpicture}
\begin{axis}[
    width=\columnwidth,
    height=5.5cm,
    xlabel={Cost per review (USD)},
    ylabel={F1 Score},
    xmin=0,xmax=0.012,
    ymin=0.28,ymax=0.40,
    grid=major,
    grid style={dashed,gray!30},
    every node near coord/.append style={font=\scriptsize,above},
]
\addplot[dashed,blue!50,thick,no markers] coordinates {(0.0009,0.326) (0.0031,0.365)};
\addplot[only marks,mark=*,mark size=3pt,blue!70,nodes near coords,
    point meta=explicit symbolic]
    coordinates {(0.0031,0.365) [Haiku]};
\addplot[only marks,mark=*,mark size=3pt,red!70,nodes near coords,
    point meta=explicit symbolic]
    coordinates {(0.0101,0.343) [Sonnet]};
\addplot[only marks,mark=*,mark size=3pt,green!60!black,nodes near coords,
    point meta=explicit symbolic]
    coordinates {(0.0009,0.326) [GPT-5.4m]};
\addplot[only marks,mark=*,mark size=3pt,orange!80,nodes near coords,
    point meta=explicit symbolic]
    coordinates {(0.0013,0.322) [Minimax]};
\addplot[only marks,mark=*,mark size=3pt,purple!60,nodes near coords,
    point meta=explicit symbolic]
    coordinates {(0.0014,0.310) [GLM]};
\end{axis}
\end{tikzpicture}
\caption{Cost-efficiency frontier. Haiku and GPT-5.4m define the Pareto frontier (dashed). Sonnet is dominated: lower F1 at 3.2$\times$ the cost of Haiku.}
\label{fig:pareto}
\end{figure}
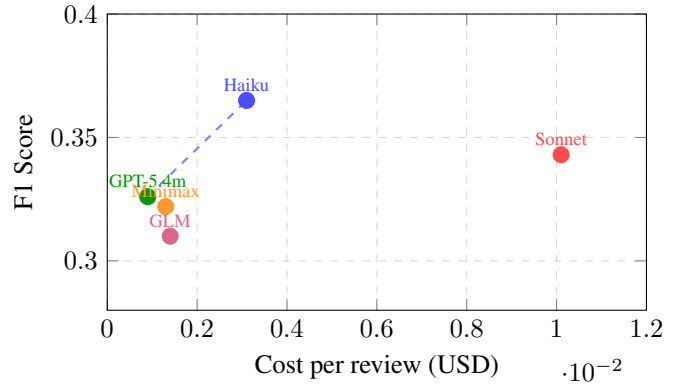

Figure~\ref{fig:pareto} plots F1 against cost per review. Two models define the Pareto frontier: GPT-5.4~mini (lowest cost) and Haiku~4.5 (highest quality). Sonnet~4.6 is \emph{dominated}: Haiku achieves higher F1 at 3.2$\times$ lower cost.

\subsection{Qualitative Assessment}

\begin{table}[h]
\centering
\small
\caption{Qualitative scores (1--5, Claude Opus judge, $n{=}150$).}
\label{tab:qual}
\begin{tabular}{lrrrr|r}
\toprule
\textbf{Model} & \textbf{Depth} & \textbf{Ctx.} & \textbf{Spec.} & \textbf{Sugg.} & \textbf{Mean} \\
\midrule
\textbf{Haiku 4.5} & \textbf{2.84} & \textbf{2.79} & \textbf{3.39} & \textbf{2.79} & \textbf{2.95} \\
GPT-5.4 mini & 2.65 & 2.54 & 3.03 & 2.44 & 2.67 \\
Sonnet 4.6 & 2.59 & 2.53 & 2.99 & 2.42 & 2.63 \\
Minimax M2.7 & 2.43 & 2.37 & 2.80 & 2.41 & 2.50 \\
GLM-5 Turbo & 2.31 & 2.29 & 2.61 & 2.26 & 2.37 \\
\bottomrule
\end{tabular}
\end{table}

Haiku scores highest on all four dimensions. The widest margin is on specificity (3.39 vs.\ Sonnet's 2.99, +13.4\%), indicating more concrete and actionable findings. Sonnet places third, behind both Haiku and GPT-5.4~mini.

\section{Discussion}

\subsection{Why Does Haiku Outperform Sonnet?}

The output behavior data (Table~\ref{tab:behavior}) suggests several contributing factors:

\paragraph{1. Calibration threshold.} Code review requires balancing comprehensiveness with restraint. Sonnet's deeper reasoning may set an overly conservative threshold; it produces 27\% fewer findings than Haiku (135 vs.\ 186) and 30\% fewer output tokens, despite greater capacity.

\paragraph{2. Pattern recognition sufficiency.} The bugs in our benchmark (SQL injection, null dereferences, missing error handling) are well-represented in training data. For these patterns, additional reasoning depth yields diminishing returns.

\paragraph{3. Output format compliance.} The prompt requires structured JSON with 8+ fields per finding. Smaller instruction-tuned models may comply more consistently, producing more complete and parseable output.

\paragraph{4. Real-PR generation gap.} On real PRs, Haiku averages 2.0 findings per sample versus Sonnet's 1.1. The higher attempt rate translates directly to recall gains. This is consistent with \citet{hsieh2023distilling}, who showed model size advantages are task-dependent.

\subsection{The Synthetic Evaluation Trap}

The 92\% F1 degradation on real-only samples (Table~\ref{tab:gap_isolated}) deserves scrutiny. Synthetic mutations are localized (median 5 lines), familiar (well-studied patterns), and unambiguous (single injected bug). Real bugs share none of these properties. Evaluations on synthetic benchmarks alone~\citep{li2022automating, tufano2022using, tufano2021towards} may overstate capability by up to 12$\times$ ($0.847 / 0.066$). Future evaluations should include real PRs, or at minimum report synthetic and real results separately.

\subsection{The Performance Blind Spot}

All five models fail on performance bugs (0\% recall for four, 4.5\% for one). Detecting N+1 queries and unbounded queries requires reasoning about execution context, data scale, and architectural role, information that PR diffs simply do not contain. Static analysis (AST pattern matching, data flow analysis) remains necessary for this category.

\subsection{Implications for Practitioners}

\begin{enumerate}[nosep,leftmargin=*]
\item \textbf{Use smaller models.} Haiku-class models match or exceed Sonnet-class at 3$\times$ lower cost.
\item \textbf{Preprocess large diffs.} The 15$\times$ F1 degradation on large diffs (Figure~\ref{fig:diffsize}) is the most actionable finding.
\item \textbf{Supplement with deterministic rules.} Performance bugs and architectural issues need AST-based detection.
\end{enumerate}

\subsection{Limitations}

\begin{enumerate}[nosep,leftmargin=*]
\item \textbf{Judge bias.} Claude Opus judged all models, including Anthropic's own. Qualitative scoring may favor Anthropic's output style.
\item \textbf{Ground truth quality.} Real PR annotations are auto-extracted, not human-labeled. Some may capture refactoring alongside bugs.
\item \textbf{Single prompt.} Provider-specific optimization may change rankings.
\item \textbf{No fine-tuning.} All models evaluated in general-purpose configuration.
\item \textbf{Temporal validity.} Results reflect model versions as of March 2026.
\item \textbf{Sample size.} 150 samples (50 real) identifies large effects but not small differences with statistical significance.
\end{enumerate}

\subsection{Data and Code Availability}

The evaluation framework, synthetic mutation engine, and raw results are available in the \texttt{vibeops-mcp/evals/} directory of the VibeOps repository. Synthetic samples are reproducible from seed~42. Real PR samples are derived from public repositories via provided mining scripts. We plan to release the dataset on HuggingFace Datasets following human validation of the real PR annotations.

\section{Conclusion}

Five LLMs were evaluated on automated code review across 150 samples. Four conclusions stand out:

\begin{enumerate}[nosep,leftmargin=*]
\item \textbf{Bigger isn't always better.} Haiku~4.5 outperforms Sonnet~4.6 on F1, recall, and all qualitative dimensions at 3.2$\times$ lower cost, across three conditions and confirmed on the Martian benchmark.
\item \textbf{Synthetic benchmarks are dangerously optimistic.} F1~$=$~0.066 on real PRs vs.\ 0.847 on synthetic, a 92\% drop.
\item \textbf{Diff size is the key bottleneck.} F1 drops 15$\times$ between small and large diffs.
\item \textbf{Performance detection is beyond current LLMs.} Near-zero recall on N+1 queries and unbounded queries.
\end{enumerate}

For production systems, Haiku~4.5 is the recommended primary model. Deterministic rules and diff preprocessing address the categories where LLMs fail. Ensembling multiple models does not improve results.

\section*{Acknowledgments}
This work was conducted as part of the VibeOps project.

\bibliography{references}

\end{document}